\documentclass[ngerman, italian, english, UKenglish]{scrreprt}
\usepackage[latin9]{inputenc}
\usepackage{geometry}
\usepackage{fancyhdr}
\usepackage{float}
\usepackage{units}
\usepackage{amsmath}
\usepackage{amssymb}
\usepackage{graphicx}
\usepackage{subfigure} 
\usepackage{caption}
\usepackage{textcomp}
\usepackage{setspace}
\usepackage{tikz}

\usepackage{hyperref}

\makeatletter

\usepackage{enumitem}

\usepackage{tikz}
\usetikzlibrary{scopes}
\usepackage{marvosym}
\usetikzlibrary{shapes,snakes}
\usepackage{fancyhdr}
\usepackage{chngcntr}
\usepackage{wallpaper}
\usepackage{multicol}
\usepackage{upgreek}
\usepackage{url}
\usepackage{pdflscape}
\usepackage[exponent-product = \cdot, output-decimal-marker={,},load-configurations=abbreviations]{siunitx}
\makeatother

\usepackage{babel}

\newcommand{\be}{\begin{equation}}
\newcommand{\ee}{\end{equation}}
\newcommand{\bea}{\begin{eqnarray}}
\newcommand{\eea}{\end{eqnarray}}

\counterwithout{equation}{chapter}
\counterwithout{table}{chapter}
\counterwithout{figure}{chapter}

\begin{document}
%%%===================================================================================%%%
\setcounter{figure}{0}
\setcounter{equation}{0}
\setcounter{table}{0}
\setcounter{page}{1}
%%%===================================================================================%%%
\newpage
\section*{Light, delayed: The Shapiro Effect and the Newtonian Limit}  %Title
\addcontentsline{toc}{section}{}

%Author's name, Affiliation
Markus P\"ossel, Haus der Astronomie and Max Planck Institute for Astronomy
\bigskip{}

\noindent \emph{} %Abstract
The Shapiro effect, also known as the gravitational time delay, is close kin to the gravitational deflection of light that was the central topic of our Summer School.\footnote{This text has been published in K.-H. Lotze \& Silvia Simionato (eds.), {\em Proceedings of the Heraeus Summer School ``Astronomy from 4 Perspectives: Thinking Gravitational Lensing for Teaching''} (Jena, 2--7 Sep 2019), pp. 42--54.} It is also an interesting test bed for exploring a topic that provides the foundations for most of the calculations we have done in this school, yet is highly complex when treated more rigorously: the question of the Newtonian limit, and of the post-Newtonian corrections that must be applied to include the leading-order effects of general relativity. This contribution discusses simplified derivations for the gravitational redshift and the Shapiro effect, as well as astrophysical situations in which the Shapiro effect can be measured.
\setlength{\columnsep}{0.75cm}

\begin{multicols}{2}

\subsubsection*{1 Introduction} %Article
Physical theories stand and fall with having their predictions tested and confirmed. For Einstein's theory of general relativity, there are four so-called classical tests: four distinct physical effects that are amenable to observational or experimental examination. From a modern point of view, it could be argued that gravitational waves, first predicted in 1916 and first detected directly in late 2015 constitute an additional test, but somehow, gravitational waves are not usually counted in that category --- probably because the detections themselves are always also astronomical observations, and never merely tests of a specific physical effect.

The first classical test was, at least in its original incarnation, not so much a prediction as a postdiction: The anomalous precession of the planet Mercury --- a minute divergence of the planet's orbit from the Newtonian prediction --- had been observed as early as 1859. Einstein made use of this anomaly when he formulated his theory of gravitation, using the successful derivation of the anomalous precession as a touchstone to identify the correct field equations among several competing candidate theories. Much later, observations involving binary pulsars would successfully test general-relativistic predictions of anomalous precession in a significantly stronger gravitational field.

Another classical test is the prediction (and confirmation) that is the focus of our teacher training: the gravitational deflection of light near massive bodies, correctly predicted by Einstein in 1915, and first confirmed during a solar eclipse in 1919, which led to a substantial widening of the acceptance of general relativity, and made Einstein famous --- and which is the subject of several other contributions in this volume.

The third classical test was Einstein's prediction of the gravitational redshift, as early as 1907, eight years before he had found the final form for general relativity--- the change of the wavelength of light as the light falls down into, or climbs out of, a gravitational well. As several authors have pointed out, this is not, strictly speaking, a test of general relativity as such, but of the Einstein equivalence principle, one of the basic assumptions on which general relativity is built. On the plus side, this makes it straightforward to derive the formula for the gravitational redshift from the equivalence principle, even in a high school setting. We will reproduce one such argument in section 2. Experimentally, the redshift was only confirmed only as late as 1960, with the Pound-Rebka experiment.

The fourth test, the gravitational time delay, is different. It was proposed much later than the first three tests, namely in the early 1960s, by Irwin I. Shapiro, and confirmed a few years later. The context suggests an interesting example for the interaction between basic physics and technology. After all, for ordinary astronomical observations, propagation times are irrelevant --- we do not know exactly when a particular photon was emitted, hence can deduce little from its arrival time. But Shapiro worked in the new field of radar astronomy, and radar works by sending electromagnetic waves out, receiving reflected portions of that light, and timing the travel time to deduce the distance of the reflecting object. In that context, propagation times and possible time delays become observable. By Shapiro's own account \cite{Shapiro2019}, the idea came to him in 1961 or 1962 after a talk given by George W. Stroke (best known as one of the pioneers of holography) at MIT, on measuring the speed of light. If you send a radar signal to a planet near superior conjunction, when the planet, the Sun and the Earth are almost lined up, in that order, the signal will pass close to the Sun; by measuring echo times, you can at least determine the total travel time of the signal, there and back again.

\subsubsection*{2 Limits of general relativity} 

Shapiro's eventual calculations made use of a weak-gravity approximation of Einstein's theory. This ties in with the more general question of limiting cases of general relativity --- a topic that is relevant for teaching general relativity in an undergraduate or high school setting, as well.

One limit is that of slow motion, $v\ll c$. From gravity-free physics, we know this as the limit where the laws of special relativity can be approximated by the laws of classical mechanics. When gravity comes into play, then for consistency, we need to avoid situations in which objects are accelerated gravitationally to high speeds, which means avoiding compact masses. Given that, according to classical mechanics, the speed reached by an object falling from infinity onto the surface of a spherical body with mass $M$ and radius $r$ is $\sqrt{2GM/r}$, we can reformulate this additional condition as $c\gg v=\sqrt{2GM/d}$ for all masses $M$ and length scales $d$ involved, or equivalently $d\gg 2GM/c^2$.

\begin{figure}[H]
\begin{center}
\includegraphics[height=0.5\columnwidth]{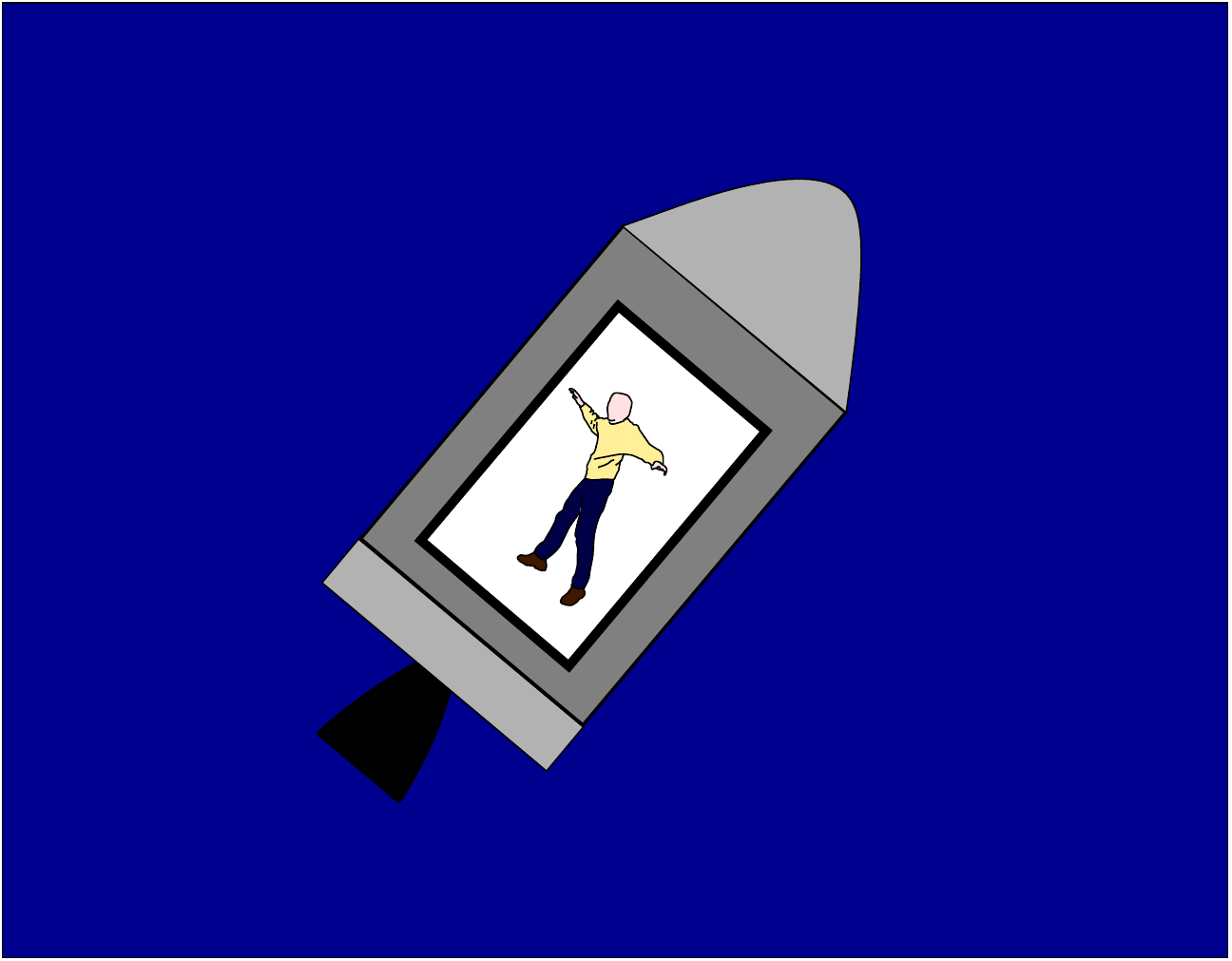}\\[1em]
\includegraphics[height=0.5\columnwidth]{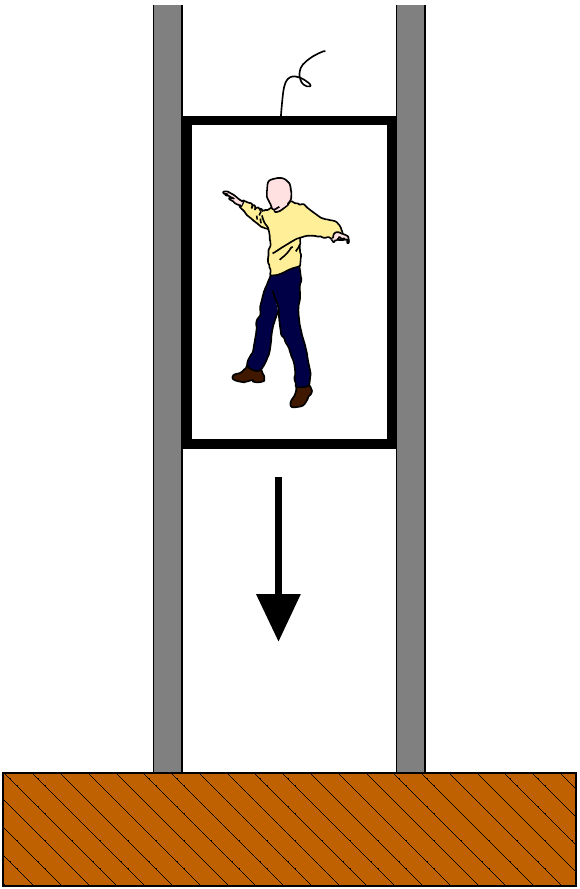}
\caption{\small Experimenters in a small cabin, with limited time for their experiments, cannot distinguish whether their cabin is in virtually gravity-free deep space or in free fall in a gravitational field}
\label{Fig:ElevatorRocketEEP}
\end{center}
\end{figure}

Another limit is the one that we reach by going into free fall, and then only looking at a small spacetime region, namely special relativity. By the Einstein equivalence principle, if we go into free fall, the only gravitational influence that remains are tidal forces. Within a sufficiently small spacetime region, tidal forces can be ignored, and the laws of physics can be approximated by the laws of special relativity. Within a suitably small laboratory, over a suitably small period of time allocated for experimentation, we cannot distinguish between the situation where the laboratory is drifting in deep space, far from all major sources of gravity (top panel of Fig.~\ref{Fig:ElevatorRocketEEP}) or is in free fall in a gravitational field (bottom panel).
\begin{figure*}[t]
\begin{center}
\includegraphics[width=0.7\textwidth]{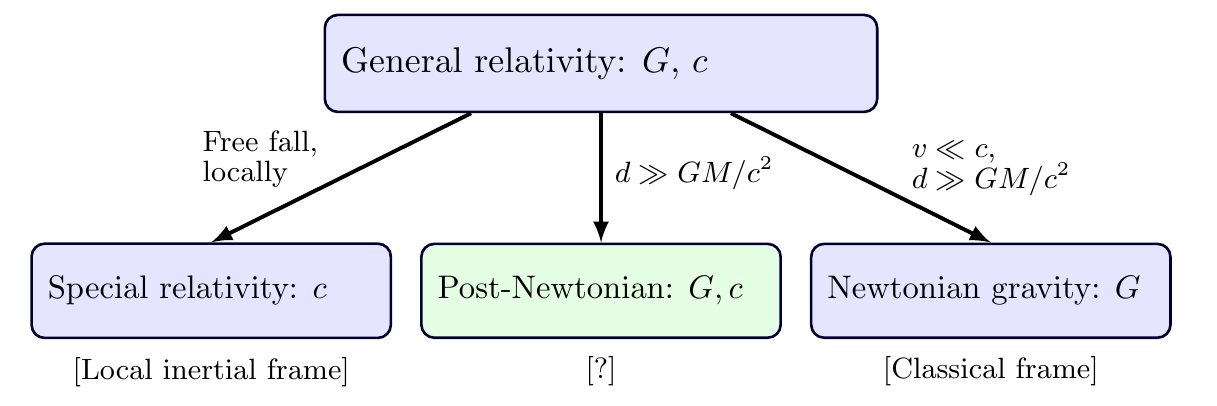}
\caption{\small Different limits of general relativity}
\label{Fig:GRLimits}
\end{center}
\end{figure*}

As mentioned in the introduction, this limiting case can be utilized to derive the formula for the gravitational redshift, as we will do in section 3. More generally speaking, the Einstein equivalence principle enables us to derive specific (possibly approximated) results without the need for the full formalism of general relativity.

The most interesting limit, however, is the one that goes to neither extreme: Where gravitational effects are small, but significant, yet where we do not want to restrict ourselves to situations in which $v\ll c$. This is known as the {\em post-Newtonian regime}, and if you had hoped that it is somehow simpler to understand than the full theory of general relativity, you would be sorely disappointed. Taking this limit in a suitably rigorous way is quite a challenge. While the basics were worked out by Vladimir Fock in the USSR and Subrahmanyan Chandrasekhar in the US in the 1960s, the last conceptual holes were only filled in in the 1990s \cite{Will2011}, and the uncertainties of the early derivations are linked directly to the question whether or not gravitational waves are real, and whether or not their emission is described by the so-called quadrupole formula \cite{Kennefick2007}.

But even so, there are simplified derivations of relativistic effects that are based on somehow linking a sort-of-post-Newtonian limit, where there is gravity, with studies of the relativistic propagation of light. If post-Newtonian relativity is so hard to get right, why do those simplified derivations work? Sure, in some cases they are missing a factor 2, but still. Stealing the quotation from Will in \cite{Will2011}, who in turn was stealing it from Eugene Wigner, we can ask: Why are simple sort-of-Newtonian models so unreasonably effective?

The Shapiro effect, as we will see, is a nice example that indicates that the simplest sort-of-Newtonian calculations are in fact not representative. Each only works because of additional, very special circumstances. Take what are by far the most prominent such derivations: Why can we describe cosmological models in an almost purely Newtonian way, deriving the dependence of the cosmic scale factor evolution $a(t)$ in terms of the energy density of the universe from Newton's law of gravity (with minor modifications for pressure terms) \cite{Poessel2017}? Only because of the homogeneity and isotropy of expansion, which allows us to calculate the dynamics of the scale factor at arbitrarily small scales, corresponding to arbitrarily slow motion of macroscopic test particles in that expanding universe, corresponding to an arbitrarily good fit of the Newtonian approximation. 

And why do we get a sensible result, off by a constant factor of two, from simple Newtonian calculations of light deflection (as those that are presented in the contribution of Karl-Heinz Lotze to these proceedings, on p. \ref{LotzeModels}ff.)? A closer look --- more precisely: an attempt to calculate the Shapiro time delay using the same approximations --- shows that the calculations get the timing of the propagation of light near a mass completely wrong, not only quantitatively but qualitatively: photons propagating as Newtonian test particles get faster as they approach the central mass, while the Shapiro time delay results from the coordinate speeds of light getting slower near the mass. Why, then, do these approximations give a sensible result for the deflection angle? Evidently because changes in direction are encoded in the ratios of velocity components (e.g. $v_y/v_x$ if we are describing light deflection in the xy plane), and overall changes in propagation speed do not affect such ratios.

A schematic overview of the different limits is given in Fig.~\ref{Fig:GRLimits}. Considering the above considerations, we might ask: Given that reasonably rigorous post-Newtonian calculations are hard, and that the simplified derivations are limited, what strategy {\em can} we trust in deriving general-relativistic effects without the full machinery of differential geometry?

\subsubsection{3 Gravitational redshift}
\label{Sec:GravitationalRedshift}

When it comes to statements about general-relativistic spacetimes without employing the full formalism, we are on comparatively safe ground whenever we employ the principle of equivalence --- for any effect that can be described in an infinitesimally small spacetime region, where tidal effects are by definition infinitesimally small, we need only go into free fall and apply the laws of special-relativistic physics. 

In cosmology, to give an example, this strategy allows us to calculate the exact formula for the propagation of light in a flat universe, and the cosmological redshift --- simply by using the fact that galaxies whose motion is determined by cosmic expansion only (``galaxies that are in the Hubble flow'') are in free fall \cite[section 6.3]{Poessel2017}, and without the need to understand or evaluate the spacetime metric.

In the situation we are interested in for the Shapiro effect, namely the propagation of light near a mass, introducing frames in free fall as prerequisite for applying the equivalence principles requires that we already have a description of gravity at hand. The only description of gravity we {\em do} have at hand at this point is, of course, Newtonian gravity. This suggests a natural way towards heuristic post-Newtonian arguments in a weak gravity situation: As a first step, we choose a coordinate system in which classical mechanics and Newton's law of gravity provide an adequate approximate description of how objects are accelerated. By its definition, we can use that coordinate system to describe the gravitational experienced by objects moving slowly, that is, with $v\ll c$. But classical physics can tell us nothing about relativistic effects, like varying clock rates or the propagation of light. That is where local free-falling frames come into play: In those frames, at least locally, special relativity allows us to describe light propagation and time dilation effects. Combining physics in those local free-falling frames with the global Newtonian coordinates governing the frames' free fall, we obtain our results.

By construction, it is also straightforward to say which element is missing in our description. Our recipe cannot include any relativistic tidal effects, that is, curvature effects. In our Newtonian coordinates, space is Euclidean, and our recipe, as we have stated it here, does not provide corrections to that spatial geometry. That is a limitation of this approximation, and when comparing results obtained in this way for light deflection and for the Shapiro time delay, the discrepancy, which amounts to a factor 2 in both cases, can indeed be traced to the neglected contributions of spatial curvature.

\begin{figure}
\begin{center}
\includegraphics[width=0.7\columnwidth]{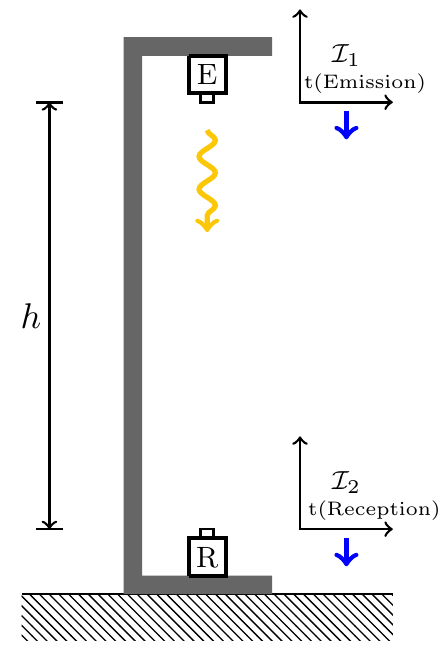}
\caption{\small Setup of free-fall coordinate systems to derive the gravitational redshift}
\label{Fig:GravitationalRedshift}
\end{center}
\end{figure}
As an example, consider the gravitational redshift, for which we will give a not-quite-rigorous derivation \cite[for a similar derivation]{Schroeter2002}. The setup can be seen in Fig.~\ref{Fig:GravitationalRedshift}: We want to examine light travelling in a homogeneous gravitational field, from an emitter E a bit higher up to a receiver R lower down. We assume the gravitational acceleration to be $g$. In order to calculate how light propagates in this scenario, we introduce two free-falling systems. The system ${\cal I}_1$ is at rest at the location of E at the exact event when the light is emitted. Assume that the wavelength of the light is so short that it can be measured locally, directly at the emission event. For this measurement, an observer at rest in ${\cal I}_1$ will obtain the same result as an observer at rest relative to the emitter, since at that event, both systems are at rest relative to each other.

Let us assume that tidal effects are so small that we can neglect them over the distance $h$ between emitter and receiver. In that case, we can extend ${\cal I}_1$ downwards to the receiver. At the moment the light arrives at the receiver, which (up to higher-order corrections) happens a time interval
\begin{equation}
\Delta t = h/c
\end{equation}
after the emission event, system 1 has gathered some speed and is now moving at
\begin{equation}
v = g\cdot\Delta t=\frac{gh}{c}.
\end{equation}
In ${\cal I}_1$, the light experiences no wavelength shift. After all, in that system, the light is propagating freely, as in special relativity. But at the moment of reception, when the second free-fall system, ${\cal I}_2$, is at rest relative to the receiver, ${\cal I}_1$ has speed $v$ relative to ${\cal I}_2$. By the classical Doppler effect (which is a valid approximation in situations when the speed is small, and thus when $\Delta t$ is small), this speed corresponds to a wavelength shift 
\begin{equation}
z \equiv \frac{\lambda_R - \lambda_E}{\lambda_E}
\label{DopplerFactor}
\end{equation}
from the emitted wavelength $\lambda_E$ of
\begin{equation}
z = -\frac{gh}{c^2} = -\frac{\Delta\Phi}{c^2},
\end{equation}
where $g\sim 9,81 \; \mbox{m/s}^2$ is the gravitational acceleration and
$\Delta\Phi = \Phi(r_E)-\Phi(r_R)$ the potential difference.

\begin{figure}[H]
\begin{center}
\includegraphics[width=\columnwidth]{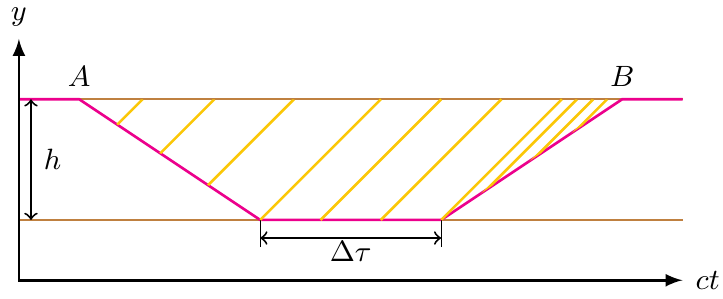}
\caption{\small Clock transported to a lower location in a gravitational field and back up again, for comparison with a clock that has remained in the original position throughout}
\label{Fig:ClockComparison}
\end{center}
\end{figure}

The gravitational redshift can be understood in terms of different ``tick rates'' of local clocks at rest in different positions in the gravitational field. Consider a coordinate system in which the gravitational potential is not time dependent, and consider two clocks at rest in that coordinate system. In such a situation, it will always take light the same coordinate time to propagate from one clock to the other. After all, the situation is, per definition, invariant under translations in time --- neither the position of the two clocks nor the gravitational potential at each point of the light's depends on the time coordinate. If two light pulses are sent from the one clock to the other, a coordinate time interval $\mathrm{d} t$ apart, then they will arrive at the second clock the same coordinate time interval $\mathrm{d} t$ apart. If these coordinate time intervals correspond to different proper time intervals --- that is, to different time intervals as measured by the clocks themselves, using physical units defined in the same way --- that is a valid reason to say that the clocks are running at different speeds. If you are skeptical since that statement is so far only based on indirect comparison via light signal, consider the situation sketched in Fig.~\ref{Fig:ClockComparison}. There, a clock is transported from its original location to a location lower in a gravitational potential, left there until $\Delta\tau$ on that clock has elapsed, and then back for comparison purposes. 
If we repeat the experiment several times, always effecting the transport in the same way, but varying the duration of $\Delta\tau$, we will find that the time $\Delta\tau_U$ that has elapsed between the events $A$ and $B$ on the clock that stays at constant height $h$ (brown worldline) is related to $\Delta\tau$ as 
\begin{equation}
\Delta\tau_U =b + \frac{\Delta\tau}{1-\frac{gh}{c^2}},
\end{equation}
where $b$ stands for the proper time that elapses on the travelling clock during the (identical) transport phases. This is consistent with the claim that the clock is running more slowly when it is lower down in the gravitational field.

While this derivation yielded the correct result, we can, if we take another critical look at it, find some weak spots. We have taken light to propagate at the speed $c$ not only in the inertial systems, but also in the Newtonian coordinate systems. And we have assumed that the coordinate speed $v$ is also the speed in the inertial system ${\cal I}_2$. If we now redo the derivation in a more careful way, we will automatically find the key ingredient for the Shapiro time delay.

\subsubsection*{4 Shapiro time delay}

The analysis of gravitational redshift in the previous section has given us valuable clues as to which elements we need to include in order to analyse the propagation of light in a gravitational field. We cannot use the purely Newtonian approximation since we want to describe light propagation, $v=c$. This suggests making use of the principle of equivalence. As we shall see below, that strategy allows us to at least capture part of what is going on in this situation. Assume that we are dealing with a radially symmetric gravitational potential $\Phi(r)$. Then we can argue as follows \cite{Poessel2019}:

First, we again choose coordinates $t$ and $r$ in which the laws of classical mechanics, and of Newtonian gravity, are approximately valid. We have learned from section 3 that, in this case, we need to distinguish between coordinate time intervals $\mathrm{d} t$ and proper time intervals $\mathrm{d}\tau$ as shown by clocks in our scenario, since the two are not necessarily the same. Let us assume explicitly that the geometry of space is Euclidean (even though that means we will necessarily miss out on at least some of the general-relativistic effects). 

By placing a momentarily co-moving, free-falling frame next to an observer, we can deduce that for an observer at rest at constant radius coordinate value $r$, light propagates at the same speed in all directions, and with the speed $c$ as measured by the proper time $\tau$ of the local clock. This means we have an $r$-dependent coordinate speed of light,
\begin{equation}
c(r) = \left|\frac{\mathrm{d} \vec{x}}{\mathrm{d} t}\right| =  \left|\frac{\mathrm{d} \vec{x}}{\mathrm{d} \tau}\right|\cdot\frac{\mathrm{d}\tau}{\mathrm{d} t} = c\cdot\frac{\mathrm{d}\tau}{\mathrm{d} t}.
\label{cOfR}
\end{equation}
In order to find the dependence of $\tau$ on $t$, we take similar steps as in our gravitational redshift calculations in section 3, but this time we are more careful about which part of the argument involves the Newtonian coordinates and which involve proper time. For instance, the inertial system ${\cal I}_1$ that is initially at rest relative to our clocks positioned at constant radius value $r+\mathrm{d} r$ will pick up a bit of coordinate speed
\begin{equation}
\mathrm{d}v = -\frac{\mathrm{d}\Phi(r)}{c(r)}
\end{equation}
by the time the light signal reaches the observer at constant radius value $r$. Making use of (\ref{cOfR}) in order to calculate the corresponding speed $\mathrm{d}v_I$, as measured by the inertial observer in ${\cal I}_2$ at the moment that observer is momentarily at rest relative to the clock at constant radius value $r$, noting that the relevant time coordinate in ${\cal I}_2$ corresponds not to the Newtonian time coordinate $t$, but to the local proper time $\tau$, we find
\begin{equation}
\mathrm{d} v_I = -c\frac{\mathrm{d}\Phi(r)}{c(r)^2}.
\end{equation}
In consequence, the Doppler factor (\ref{DopplerFactor}) in this case is 
\begin{equation}
z=\frac{\mathrm{d} v_I}{c} = -\frac{\mathrm{d}\Phi(r)}{c(r)^2}.
\end{equation}
But as we have seen, that Doppler shift also determines the relation between proper times at different radius values and the coordinate time $t$, and thus the relation between proper time intervals at different radius values. In our case, we have
\begin{equation}
\frac{\mathrm{d}\tau(r+\mathrm{d} r)}{\mathrm{d}\tau(r)} = \frac{1}{1+z}\approx 1+\frac{\mathrm{d}\Phi(r)}{c(r)^2}.
\end{equation}
This allows us to re-write 
\begin{eqnarray}
\frac{\mathrm{d} c(r)}{\mathrm{d} r} &=&c\cdot\frac{\mathrm{d}\tau(r+\mathrm{d} r)-\mathrm{d}\tau(r)}{\mathrm{d} r\cdot\mathrm{d} t}\\[1em]
&=& \frac{1}{c(r)}\frac{\mathrm{d}\Phi}{\mathrm{d} r}.
\label{DPhiDr}
\end{eqnarray}
We are interested in the situation where the masses are all localized in a limited region of space; far away from those masses, light will propagate at the vacuum speed of light $c$, same as in special relativity, so $\lim_{r\to\infty} c(r)=c$. At such great distances, the gravitational potential will vanish, $\lim_{r\to\infty}\Phi(r)=0$. (Note that the latter is not just a matter of convention any more, as in classical mechanics --- the potential determines the ticking rates of our clocks; the limit is a consequence of our assumption that, far away from the masses, physics, including clock rates, is approximately given by the physics of gravity-free space, that is, the physics of special relativity.)

With these limiting values, we can integrate up (\ref{DPhiDr}) to yield
\begin{equation}
c(r)=  c\sqrt{
1+\frac{2\Phi(r)}{c^2}
}= c\sqrt{
1-\frac{2GM}{rc^2}
},
\label{LightSpeedOfR}
\end{equation}
where in the second step we have inserted the Newtonian potential for a spherically symmetric mass. Closer to the central mass, the coordinate speed of light $c(r)$ slows down --- that is going to be what leads to the time delay. Note that this contradicts what our expectation would have been, had we treated light like a (massless) Newtonian particle, as in the Soldner or Cavendish calculations. From that Ansatz, energy conservation would have led us to 
\begin{equation}
c(r) = c\cdot \sqrt{1+\frac{2GM}{c^2r}}\approx c\left[1+\frac{GM}{c^2r}\right],
\end{equation}
with light speeding up as it approaches the central mass (just like, say, a comet will). This leads to a similar logarithmic shape, but for a counter-factual travel time {\em reduction}, not delay.
\begin{figure}[H]
\begin{center}
\includegraphics[width=\columnwidth]{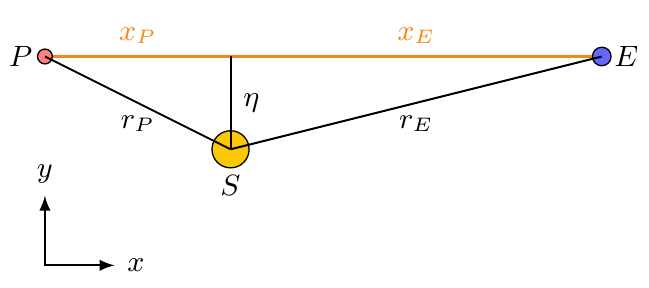}
\caption{\small Geometric setup for our time delay calculation}
\label{Fig:TimeDelaySetup}
\end{center}
\end{figure}

\begin{figure}[H]
\begin{center}
\includegraphics[width=\columnwidth]{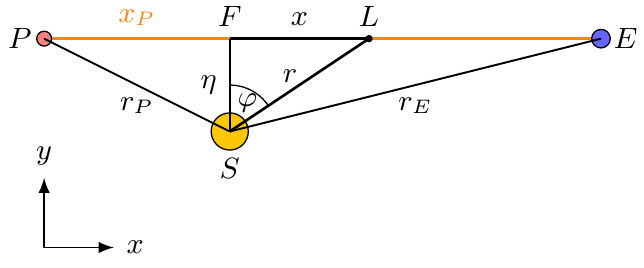}
\caption{\small Geometric setup for our time delay calculation: light trajectory}
\label{Fig:TimeDelaySetup2}
\end{center}
\end{figure}

Let us use the position-dependent speed of light (\ref{LightSpeedOfR}) to calculate the time delay for light from an object $P$ --- think of it as a planet, for now --- travelling to the Earth $E$ close to a mass $S$, such as the Sun. In this setup, whose geometry is sketched in Fig.~\ref{Fig:TimeDelaySetup}, we choose the center of the Sun as the origin of our coordinate system, place the planet at $x=x_P<0$ and $y=\eta$ and the Earth at $x=x_E>0$ and $y=\eta$. As part of our simplification, we are making the assumption that our light signal travels along a straight line. Since we are assuming Euclidean geometry, we have
\begin{equation}
r_E=\sqrt{x_E^2+\eta^2}
\end{equation}
and
\begin{equation}
r_P=\sqrt{x_P^2+\eta^2}.
\end{equation}
The $\eta$ marks the light's closest approach to the mass $S$, and thus is what is commonly called the impact parameter in scattering scenarios such as this. We describe the trajectory of the light as sketched in Fig.~\ref{Fig:TimeDelaySetup2}.
The speed at which the light is moving along the straight line between $P$ and $E$ is given by (\ref{LightSpeedOfR}), where
\begin{equation}
r=\sqrt{x^2+\eta^2}.
\end{equation}
Separating the variables $x$ and $t$, and using $1/\sqrt{1+\xi}\approx 1-\xi/2$ for small $\xi$, we have
\begin{equation}
c\,\mathrm{d} t = \left[ 1+\frac{GM}{c^2\sqrt{x^2+\eta^2}} \right]\mathrm{d} x.
\end{equation}
Looking up the integral (I myself looked on {http://wolframalpha.com}), we find
\begin{equation}
\int \frac{\mathrm{d} x}{\sqrt{x^2+\eta^2}} = \ln\left(
\sqrt{x^2+\eta^2} +x
\right)+C.
\end{equation}
For the travel time of our light signal, leaving $P$ at time $t_P$ and arriving at $E$ at the time $t_E$, this means
\begin{equation}
t_E - t_P = \frac{x_E-x_P}{c} + \frac{GM}{c^3}\ln\left[
\frac{r_E+x_E}{r_P+x_P}
\right].
\label{Eq:ShapiroDelay1}
\end{equation}
The first term on the right-hand side is the travel time for light travelling at constant speed $c$. The second term is the general-relativistic correction: the Shapiro time delay, or gravitational time delay, with its characteristic logarithmic form. 

Note that, similar to simplified Newtonian derivations of the light deflection angle, our derivation has only taken into account the effect on gravity on local clocks, but not the associated curvature of space. In consequence, again like in the case of light deflection, our result for the Shapiro time delay is too small by a factor of 2. It should be noted that even with a more rigorous derivation using the metric, it is not straightforward to get the correct result. Using the most common form of the Schwarzschild metric, one will get an unphysical extra term; only when using the so-called isotropic form of the metric describing a spherical mass will one get the same result as rigorous post-Newtonian calculations --- another indication that post-Newtonian derivations are not to be taken lightly. Simply choosing some coordinates, doing a Taylor expansion and only keeping the leading terms is not enough \cite{Poessel2019}. 

Here, we will be satisfied with having derived the correct functional form (\ref{Eq:ShapiroDelay1}), except for an overall factor of two. In the rest of this text, we will use the corrected version
\begin{equation}
\Delta t_{Sh} =  \frac{2GM}{c^3}\ln\left[
\frac{r_E+x_E}{r_P+x_P}
\right]
\label{Eq:ShapiroDelay}
\end{equation}
as we consider specific applications. There is an alternative way of writing this. Just like for the gravitational deflection of light, we can distinguish between the Newtonian result, the general-relativistic result and possible other results by introducing a parameter $\gamma$ that is zero in the Newtonian context, whereas $\gamma=1$ in general relativity. Using this parameter, we can rewrite the Shapiro delay as
\begin{equation}
\Delta t_{Sh} =  (1+\gamma)\frac{GM}{c^3}\ln\left[
\frac{r_E+x_E}{r_P+x_P}
\right],
\label{Eq:ShapiroDelayGamma}
\end{equation}
and then use observational results to constrain the value of $\gamma$. The rigorous framework for this is the {\em parametrized Post-Newtonian} (PPN) formalism which is at the heart of all modern tests of general relativity \cite{Will1986,Will2011}.

The basic time scale of the Shapiro delay is set by the factor in front of the logarithm, namely as
\begin{equation}
\frac{GM}{c^3} = (1+\gamma)\cdot 5\:\mu\mbox{s}\left(\frac{M}{M_{\odot}}\right).
\end{equation}

\subsubsection*{5 Solar System}
\begin{figure}[H]
\begin{center}
\includegraphics[width=\linewidth]{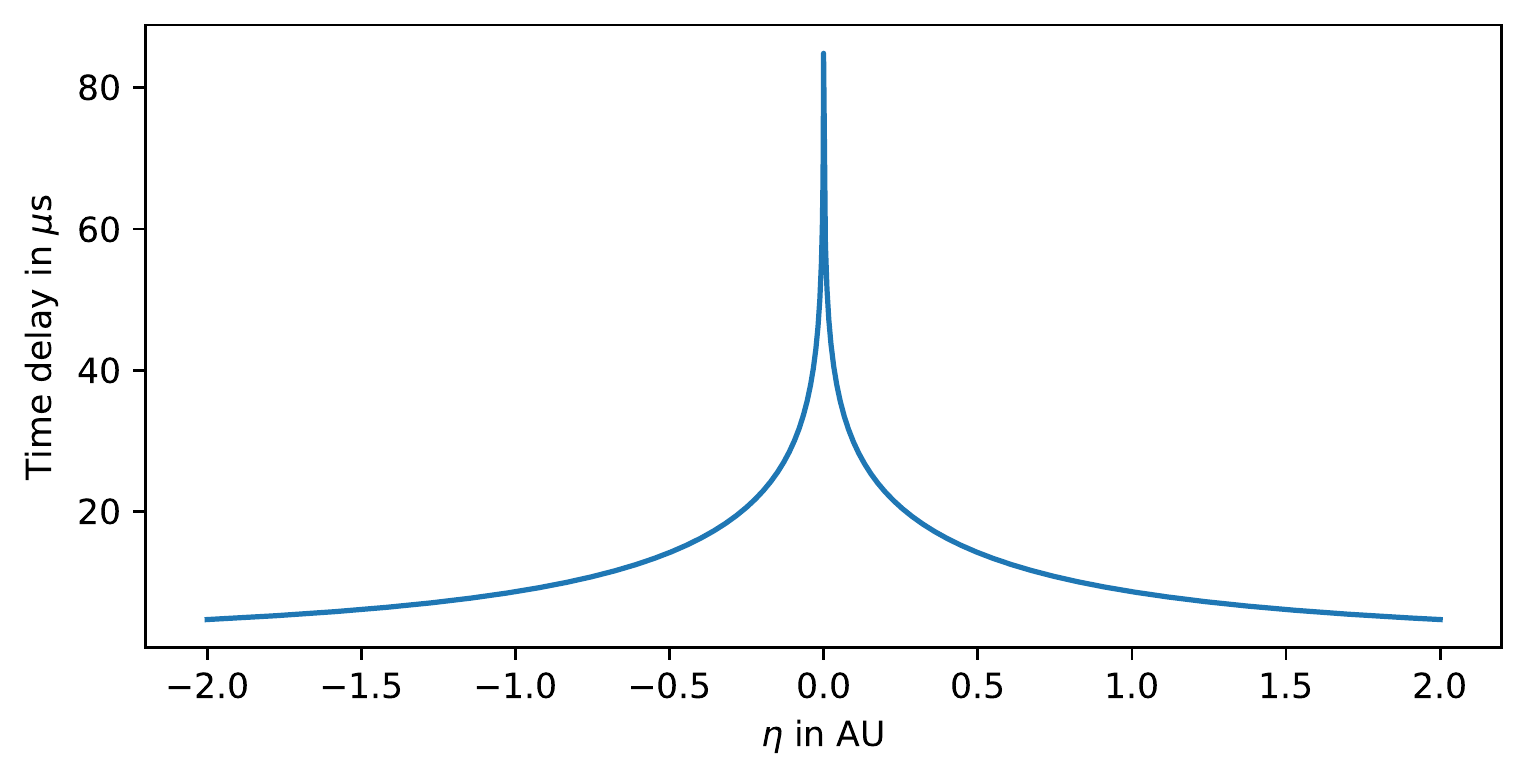}
\caption{\small Time delay for objects next to the Sun}
\label{Fig:SolSysDelay}
\end{center}
\end{figure}

\begin{figure}[H]
\begin{center}
\includegraphics[width=0.8\columnwidth]{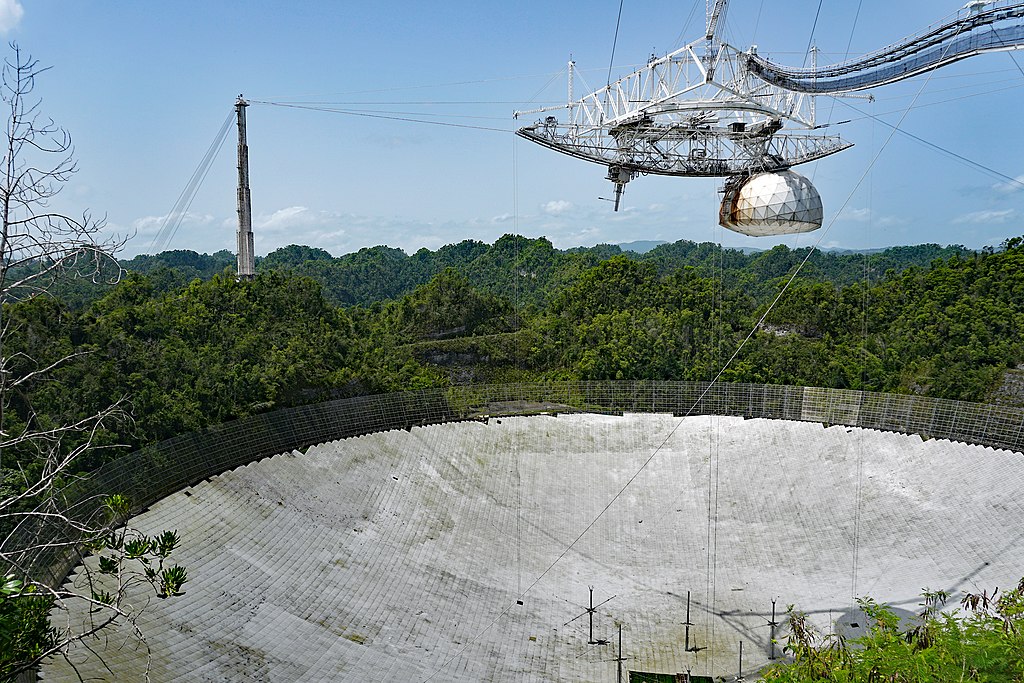}
\caption{\small The Arecibo Radio Te\-le\-scope. I\-mage by Mario Roberto Dur\'an Ortiz on Wikimedia Commons, license CC BY-SA 4.0 {\footnotesize https://creativecommons.org/licenses/by-sa/4.0/legalcode}
}
\label{Fig:Arecibo}
\end{center}
\end{figure}

The Shapiro time delay was first observed in the Solar system. To get a feeling for the scale of observations within the Solar system, consider an object for which $x_P=0$, that is, an object that is at the closest approach of its own light signals to the Sun. Now, we can plot the associated Shapiro time delay (\ref{Eq:ShapiroDelayGamma}) for $x_P=0, r_P=\eta, x_E=\sqrt{r_E^2-\eta^2}$, where $r_E$ is one astronomical unit. For $\eta$, we plot the range from zero 
The result is shown in 
Fig.~\ref{Fig:SolSysDelay}. The sharp peak with convex flanks is a very characteristic shape.
As the minimal $\eta$ value for this plot, we have taken the radius of the Sun --- no signal can pass the Sun closer than that. Also, in a typical situation, the delay plotted here, will be only part of the picture. This delay is for a signal starting at the closest distance to the Sun. For a signal starting further out, the delay will be larger --- for a signal starting at a distance of $r_E=1\:\mbox{au}$, passing the Sun, and finally reaching Earth, the delay will be twice as large as plotted here. Thus, within the Solar system, we can reasonably expect maximal delays between about $80\:\mu\mbox{s}$ and $160\:\mu\mbox{s}$. Of course, time delay is only measurable if you know how long the light was supposed to take in the first place! For this reason, Solar System measurements make use of {\em radar astronomy}, sending radio signals to another planet, detecting the reflected signal, and timing both in order to determine the round-trip travel time. There are not that many radars for use in astronomy around. Irwin Shapiro's initial measurements made use of the MIT 37m Haystack Observatory antenna as well as the Arecibo radio telescope shown in Fig.~\ref{Fig:Arecibo}.

Radar ranging plays several roles in modern astronomy. It can be used to monitor the distances to other planets with high precision, allowing both for the reconstruction of orbits and for the measurement of the astronomical unit as the basic length scale for solar system dynamics. 

Measuring the propagation of reflected radio signals sufficiently accurately to be able to detect the Shapiro time delay is a considerable challenge \cite{Shapiro1964,Shapiro1968,Shapiro1971}. First of all, mainly due to the distances involved, any signal that is sent out, reflected and sent back will be damped by a factor of about $10^{-27}$. When you send out a 100\:kW signal, what comes back will have the power of a mere $10^{-19}\:\mbox{W}$.

Then, there is the required precision needed for reconstructing what exactly is happening. In order to predict the travel time of your signal in the absence of the Shapiro delay, the geometry must be known with great precision. Only then can you distinguish between the time light would need if travelling at constant speed $c$ through Euclidean geometry --- what is called the R{\o}mer time delay, commemorating Ole R{\o}mers 17th century astronomical measurement of the speed of light --- and the travel time predicted by general relativity.

Part of the problem is that the planetary surface is not a perfect sphere, but instead a more complex landscape. On Mercury, the lowest and highest surface point have a difference of almost 10\:km in elevation; for Venus, the maximal elevation difference is about 13\:km, for Mars nearly 30\:km. A height difference of 6\:km will correspond to a travel time difference of $20\:\mu\mbox{s}$. In order to achieve the accuracy needed for the Shapiro delay measurements, the researchers first had to create detailed planetary maps (which, of course, is a useful achievement in its own right). Shapiro's team would commonly do this around inferior conjunction, when the planet was closest to Earth.

\begin{figure}[H]
\begin{center}
\includegraphics[width=0.8\columnwidth]{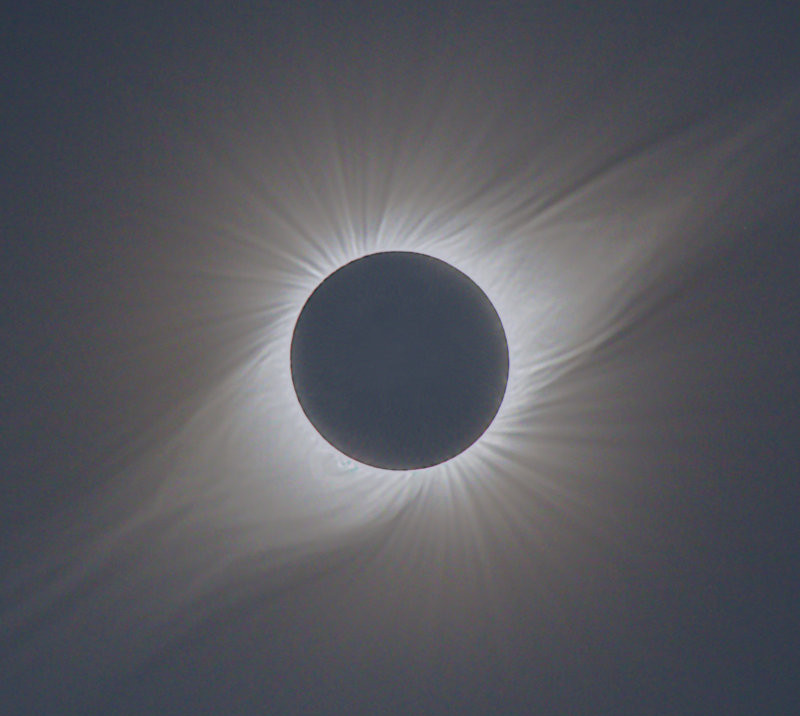}
\caption{\small Solar corona, as visible from La Silla, Chile, during the 2 July 2019 solar eclipse.
Image: M. P\"ossel, C. Liefke \& D. Els\"asser}
\label{Fig:SolarEclipse}
\end{center}
\end{figure}
There are other factors, as well. If you have ever seen an image taken during a solar eclipse, such as Fig.~\ref{Fig:SolarEclipse} from earlier this year, you know that the Sun has an extended corona. That corona is made of plasma, which in turn has an effect on the propagation of radio waves. The effect is, however, frequency-dependent, namely inversely proportional to the square of the radar frequency, and is thus a few hundred times smaller at Haystack transmitting and receiving at 8\:GHz than at Arecibo at 430\:MHz. For the 1969/1970 measurements described below, the two-way coronal delay for the Haystack measurements was estimated at about $3\:\mu\mbox{s}$, amounting to uncertainties of the order of a few percent.

How does one ``ping'' a planet? If one were to send an unchanging, continuous signal, one could not tell when the reflected radio light that reached the receiver had been sent out, and thus one could not determine the travel time. When sending out short pulses, on the other hand, in essence turning the radar transmitter off and on again, it is difficult to reach sufficiently high power. The Shapiro group solved this problem by sending a continuous signal, but (pseudo-)randomly flipping the {\em phase} of the radio wave every $60\:\mu\mbox{s}$, in a sort of cosmic Morse code. In this way, what went on the journey was a strong continuous signal, yet with a unique phase pattern imprinted that would allow the astronomers to determine exactly when a certain portion would be transmitted and when the same portion would be received roughly 30 minutes later, at what time.

On the plus side, the characteristic shape of the curve, a peak with two convex flanks as in (\ref{Fig:SolSysDelay}), which is not easily mimicked by other effects, allows for reliable fitting of the data points. 

Shapiro started his first serious attempts at measuring the gravitational time delay in the fall of 1966, after the new transmitter at the Haystack antenna had become operational --- an improvement of a factor 5 in power, with an impressive 500\:kW. His targets were two superior conjunctions of Mercury in 1967, that is, two occasions on which Mercury, the Sun and the Earth stood almost in line, in that order, so that radio waves travelling from Earth to Mercury and back would pass close to the Sun. From 12 measurements around the mid-May superior conjunction, and 7 measurements around the following superior conjunction in late August, the astronomers obtained a measurement of 
\begin{equation}
\gamma=0.8\pm 0.4,
\end{equation}
definitely favouring general relativity over the Newtonian prediction, but still with a comparatively large error \cite{Shapiro1968}. Shapiro followed this up with 
\begin{figure*}[t]
\begin{center}
\includegraphics[width=0.8\textwidth]{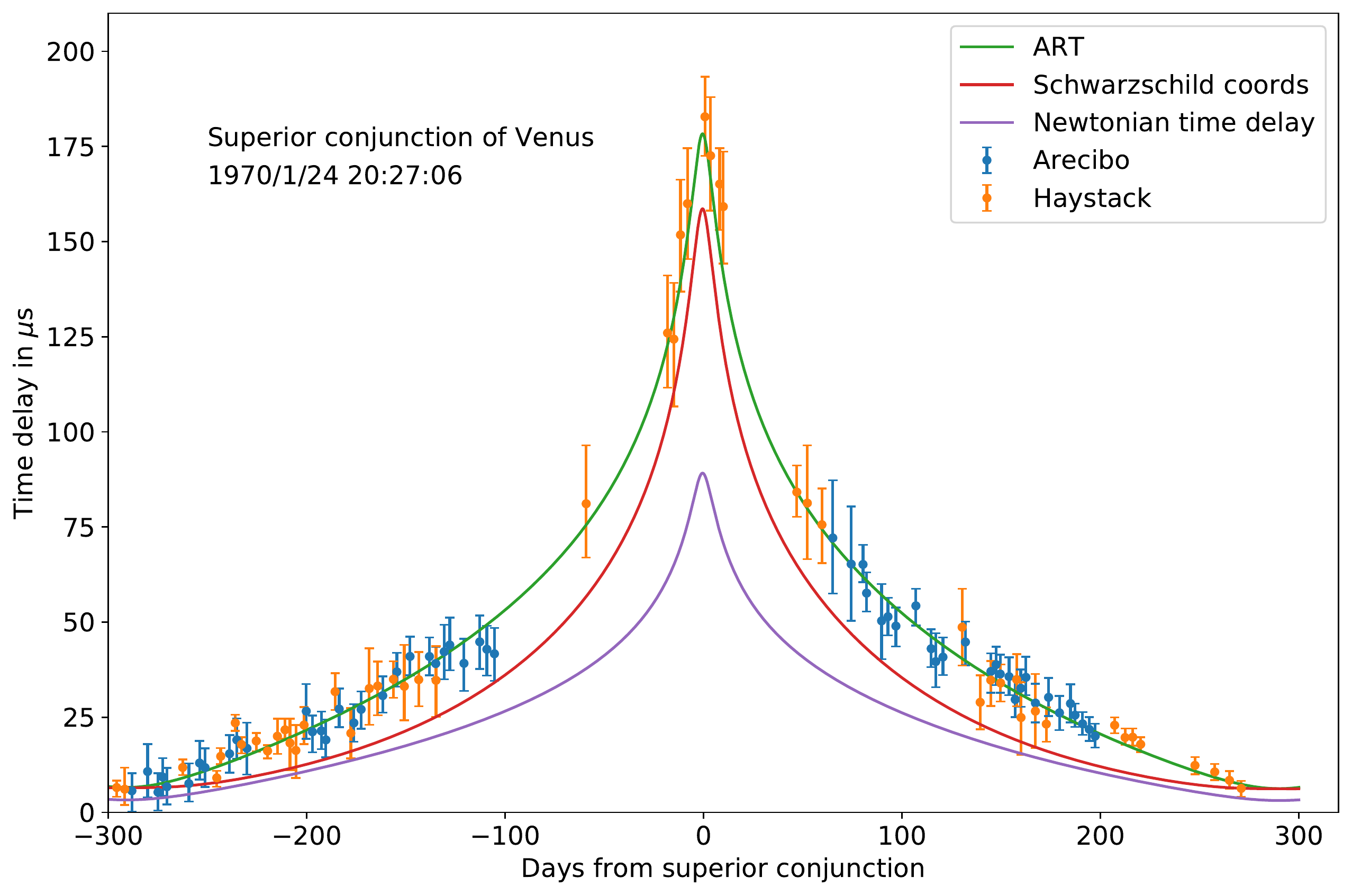}
\caption{\small Time delay of radar signals around the January 1970 superior conjunction of Venus}
\label{Fig:ShapiroVenus}
\end{center}
\end{figure*}
observations around the January 24, 1970, 20:27:06UT superior conjunction of Venus, using the Arecibo and Haystack telescopes to make radar measurements starting 300 days before the conjunction, and following up 300 days after, with Haystack covering the 120 or so days directly before and after the conjunction itself. The result is shown in Fig.~\ref{Fig:ShapiroVenus}, whose data set consists of an incomplete reproduction of that in Fig.~1 of \cite{Shapiro1971}, namely of those data points I could extract, with their error bars, from that figure. The extracted data set can be found on [{http://www.haus-der-astronomie.de/shapiro-delay}].

Three curves are shown for comparison. In green is the general-relativistic prediction from (\ref{Eq:ShapiroDelay}). In violet is the curve of the Newtonian prediction, a factor 2 below the general-relativistic one. The red curve is the prediction from the naive calculation using Schwarzschild's original form of his metric, shown here as a reminder that post-Newtonian calculations are more difficult to get right than a number of text book authors apparently think. All three curves use ephemerides generated with the ephem Python package by Brandon Rhodes \cite{Rhodes}. With this data, Shapiro and his colleagues obtained the best fit for
\begin{equation}
\gamma = 1.03\pm 0.04,
\end{equation}
a successful fit at an accuracy below the 4\% level.

Even more accurate, later measurements within our Solar system made use of space probes with transponders --- send a signal to such a spaceprobe, and its transponder will reply immediately (or at least after a well-defined short delay). Having two transmitters, one on either end, makes it considerably easier to receive the return signal. 

The first such measurements were made using the Mariner 6 and Mariner 7 space probes in 1970. The most precise such measurement was made in 2002, making use of the Cassini probe in orbit around Saturn \cite{Bertotti2003}. The result was
\begin{equation}
\gamma = 1 + (2.1\pm 2.3)\cdot 10^{-5},
\end{equation}
a confirmation of the general-relativistic prediction at the impressive level of 23 parts per million.

\subsubsection*{6 Pulsars} %As many as you think it is necessary

There is another astrophysical scenario that involves the Shapiro effect, and amounts to tests of general relativity in some of the strongest gravity environments we can study: binary pulsars. Since there is a separate talk on this by Norbert Wex, I will only recapitulate the basics. A neutron star is the compact remnant of the core of a massive star that has gone supernova. Neutron stars are only about as large as the Earth, but with a mass like that of several Suns. This makes for some of the most dense states of matter which, in this case, is mostly in the form of neutrons. Due to the conservation of angular momentum, neutron stars are typically in fast rotation (think ice-scaters doing pirouettes), and from a combination of that rotation with neutron stars' rather strong magnetic fields, neutron stars typically send beams of strong radiation into opposite directions. 

Where the magnetic poles of a neutron star are offset against the axis of rotation, those beams sweep through space like the light beams of a lighthouse. Whenever a neutron star happens to be oriented just right for one of its beams to periodically sweep over Earth, radio astronomers can observe a pulsar: every time the beam (or a part thereof) sweeps over the Earth, they register a pulse. Given that the regularity is directly linked to the neutron star's rotation, and that the star's enormous moment of inertia makes it very difficult for any physical influence to change the rotation period even by a little, the pulses we receive from a pulsar are very regular indeed --- about as regular as the most regular atomic clocks we can build.

Whenever a pulsar is part of a binary system, and when the double star partner is a compact object as well, like a White Dwarf, or another neutron star, or possibly even a black hole, that inherent regularity allows for high precision tests of general relativity. Unless we are looking onto the binary system face on, the pulsar's radio signals will occasionally pass the other star's mass more or less closely. Given that both White Dwarfs and neutron stars are comparatively compact, the general-relativistic effects in their immediate vicinity are comparatively large. As the radio signal travels from the pulsar to Earth, the proximity of the mass will influence the propagation of light --- including light deflection and a Shapiro time delay.

\begin{figure}[H]
\begin{center}
\includegraphics{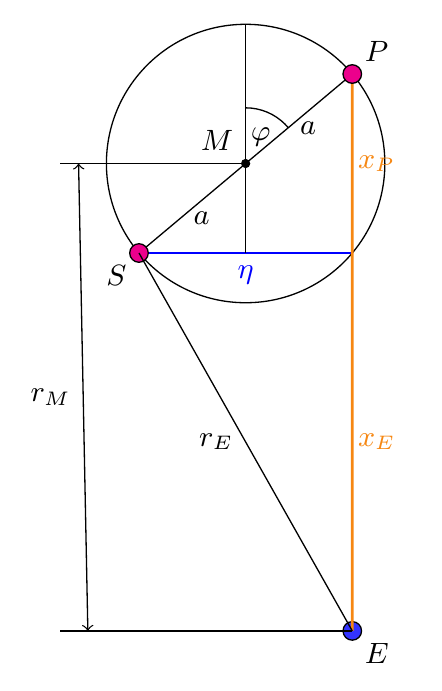}
\caption{\small Simplified geometry for a pulsar's radio waves travelling to Earth, passing the pulsar's binary partner}
\label{Fig:PulsarSetup}
\end{center}
\end{figure}
We can estimate the magnitude of the effect by looking at the simplest possible set-up, an edge-on view of an equal-mass binary pulsar, as sketched in Fig.~\ref{Fig:PulsarSetup}. In that situation,
\begin{eqnarray}
\eta &=& 2a\cdot\sin\varphi\\[0.5em]
x_P &=& -2a\cdot\cos\varphi\\[0.5em]
x_E &=& r_M +a\cdot\cos\varphi\\[0.5em]
r_P &=& 2a,
\end{eqnarray}
and we pick parameters with a sensible order of magnitude,
\begin{eqnarray}
M&=&M_{\odot}\\[0.5em]
a&=&0.01\:\mbox{AU}\\[0.5em]
r_M&=&10^9\:\mbox{AU}
\end{eqnarray}
In that situation, the basic time parameter, which in pulsar astronomy is called the Shapiro shape parameter $r$, for range, is
\begin{equation}
r = \frac{GM}{c^3} = 5\:\mu\mbox{sec}.
\end{equation}
The other Shapiro shape parameter, $s=\sin\iota$, is directly related to the system's inclination; in our simplified edge-on case, $s=1$. A simple derivation of the Shapiro time delay in terms of the phase angle $\varphi$ in a more general geometric situation can be found in \cite{Poessel2019}.

\begin{figure}[H]
\begin{center}
\includegraphics[width=\columnwidth]{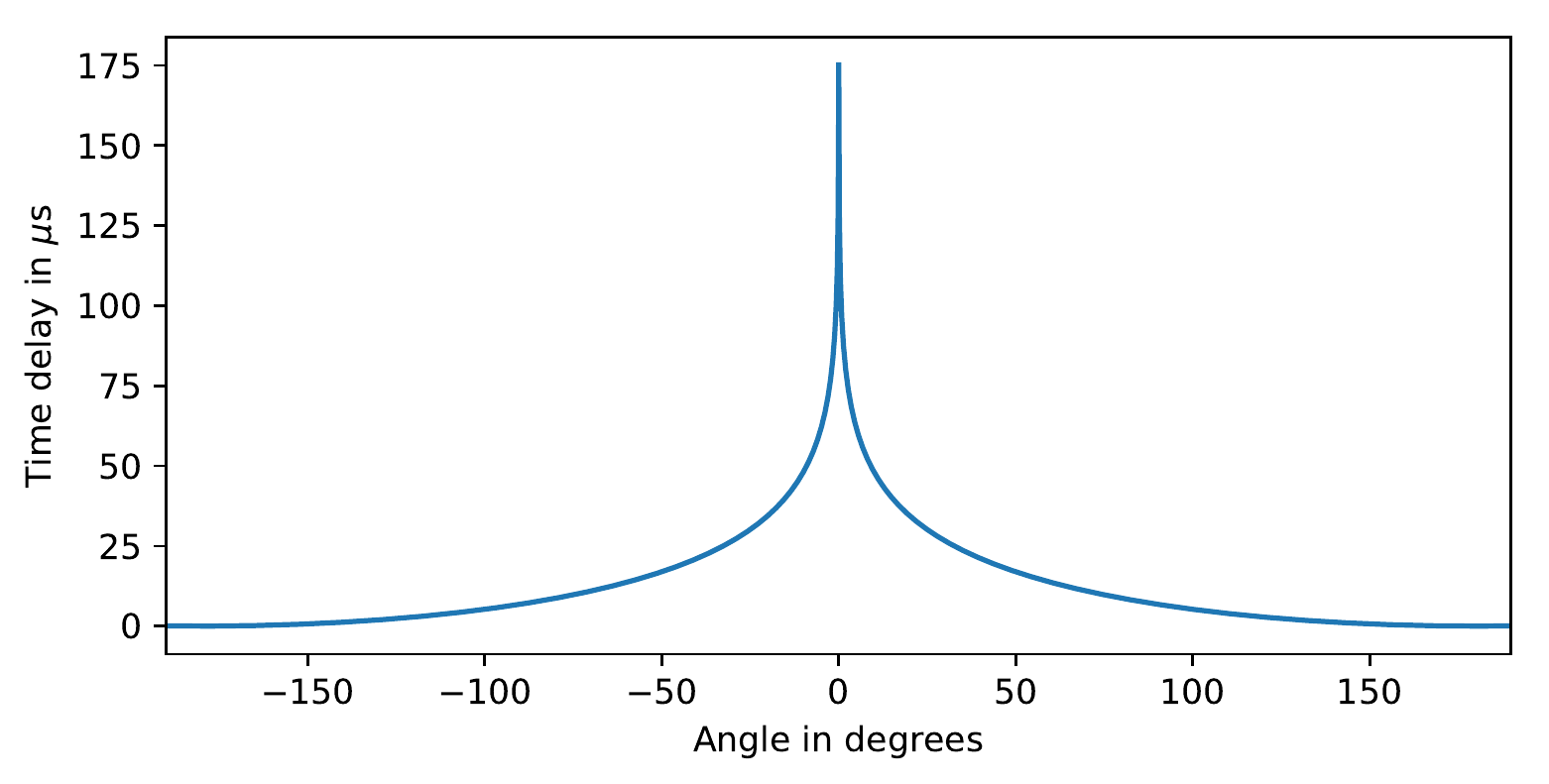}
\caption{\small Shapiro time delay for our simple edge-on neutron star binary system as a function of the phase angle $\varphi$}
\label{Fig:ShapiroPulsar}
\end{center}
\end{figure}

The result can be seen in Fig.~\ref{Fig:ShapiroPulsar}. 
With a maximum around $175\:\mu\mbox{s}$, the magnitude is similar to our solar system effects. The Shapiro shape parameters have been determined for the famous first binary pulsar PSR 1913+16  \cite{TaylorWeisberg1989}, as well as for the double pulsar PSR J0737-3039A/B \cite{Kramer2006} and for the system that has the pulsar PSR J$1614-2230$ orbiting a White dwarf \cite{Demorest2010}.

\subsubsection*{7 Conclusions}

The Shapiro effect is one of the basic general-relativistic corrections to classical gravitational physics. It is closely related to the gravitational deflection of light, and as we have seen, there are similar simplified Newtonian derivations for one and for the other, both off by a factor 2 as they neglect the curvature of space. 

Measurements of the Shapiro time delay can be used as tests for the same post-Newtonian parameter $\gamma$ that encodes the difference between the Newtonian and general-relativistic predictions for light deflection, as well. 

The gravitational time delay becomes important in two astrophysical contexts: Within the Solar system, it has been tested using radar signals reflected off other planets near and at superior conjunction, when such radar signals pass close to the Sun. The most precise versions of these measurements use space probes, with the signals in question received and sent back actively by a transponder installed on the probe.

The Shapiro time delay is also one of the effects that affect the systematic deviations from perfect regularity of the pulses we receive from pulsars that are part of a binary system. Such binary pulsars allow for some of the most stringent tests of Einstein's theory of gravity to date, and the Shapiro time delay is one of the relevant general-relativistic effects.

\end{multicols}

\subsubsection*{References}
\begingroup
\renewcommand{\chapter}[2]{}%

\endgroup
\vspace*{1ex}
\noindent Address: Haus der Astronomie, MPIA Campus, K\"onigstuhl 17, 69117 Heidelberg, Germany
 %email
E-Mail: poessel@hda-hd.de

%\ThisCenterWallPaper{1}{}
%\bigskip
%
%\noindent\hrulefill
%\clearpage


\begin{thebibliography}{99}
\small
%\bibitem{}
\bibitem{Bertotti2003} Bertotti, B., L. Iess and P. Tortora, ``A test of general relativity using radio links with the Cassini spacecraft'' {\em Nature} \textbf{425}, 374--376 (2003). doi:~\href{http://dx.doi.org/10.1038/nature01997}{10.1038/nature01997}
\bibitem{Demorest2010} Demorest, P.~B. et al., ``A two-solar-mass neutron star measured using Shapiro delay,'' {\em Nature} \textbf{467}, 1081--1083 (2010). doi:~\href{http://dx.doi.org/10.1038/nature09466}{10.1038/nature09466}
\bibitem{Kennefick2007} Kennefick, D., {\em Traveling at the Speed of Thought: Einstein and the Quest for Gravitational Waves.} Princeton University Press 2007
\bibitem{Kramer2006} Kramer, M. et al., ``Tests of General Relativity from Timing the Double Pulsar,'' {\em Science} \textbf{314}, 97--102 (2006). doi:~\href{http://dx.doi.org/10.1126/science.1132305}{10.1126/science.1132305}
%\bibitem{RennSauer2007} Renn, J. and T. Sauer 2007, ``Pathways out of Classical Physics: Einstein's Double Strategy in his Search for the Gravitational Field Equation'' in J. Renn (ed.), {\em The Genesis of General Relativity,} Vol. 1, pp. 114--312. Springer.
\bibitem{Poessel2017} P\"ossel, M., ``The expanding universe: an introduction,'' \href{http://arxiv.org/abs/1712.10315}{arXiv:1712.10315 [gr-qc]} (2017)
\bibitem{Poessel2019} P\"ossel, M., ``The Shapiro time delay and the equivalence principle,''  \href{http://arxiv.org/abs/2001.00229}{arXiv:2001.00229 [gr-qc]} (2020)
\bibitem{Rhodes} Rhodes, B., \href{https://pypi.org/project/ephem/}{https://pypi.org/project/ephem/}], last accessed 2019-12-27
\bibitem{Schiff1960}  Schiff, L,~I., ``On Experimental Tests of the General Theory of Relativity,''
{\em Am. J. Phys} \textbf{28}, 340 (1960). doi:~\href{http://dx.doi.org/10.1119/1.1935800}{10.1119/1.1935800}
% http://adsabs.harvard.edu/abs/1960AmJPh..28..340S
\bibitem{Schroeter2002} Schr\"oter, U., ``Allgemeine Relativit\"atstheorie mit den Mitteln der Schulmathematik,''  {\em Wege in der Physikdidaktik Band 5: Naturph\"anomene und Astronomie.} Ed. by Lotze, K.-H. and W. B. Schneider. Palm \& Enke, pp.\ 174--187 (2002). URL: \href{http://www.solstice.de/cms/upload/wege/band5/wege5-p2-174-187.pdf}{http://www.solstice.de/cms/upload/wege/band5/wege5-p2-174-187.pdf}
\bibitem{Shapiro1964} Shapiro, I.~I., ``Fourth Test of General Relativity,'' {\em PRL} \textbf{13}, 789 (1964). doi: ~\href{http://dx.doi.org/10.1103/PhysRevLett.13.789}{10.1103/PhysRevLett.13.789}
% http://adsabs.harvard.edu/abs/1964PhRvL..13..789S
\bibitem{Shapiro1968} Shapiro, I.~I., ``Fourth Test of General Relativity: Preliminary Results,''
{\em PRL} \textbf{20}, 1265 (1968). doi: ~\href{http://dx.doi.org/1968PhRvL..20.1265S}{1968PhRvL..20.1265S}
% http://adsabs.harvard.edu/abs/1968PhRvL..20.1265S
\bibitem{Shapiro1971} Shapiro, I.~I. et al., ``Fourth Test of General Relativity: New Radar Result,'' {\em PRL} \textbf{26}, 1132--1135 (1971). doi:~\href{http://dx.doi.org/10.1103/PhysRevLett.26.1132}{10.1103/PhysRevLett.26.1132} 
% http://adsabs.harvard.edu/abs/1971PhRvL..26.1132S
\bibitem{Shapiro2019} Shapiro, I.~I. 2020, ``The gravitational time delay,'' {\em Einstein Online}, to be published.
\bibitem{TaylorWeisberg1989} Taylor, J.~H. and J.~M. Weisberg, ``Further Experimental Tests of Relativistic Gravity Using the Binary Pulsar PSR 1913+16'' in {\em Astrophysical Journal} \textbf{345}, pp. 434--450 (1989). doi:~\href{http://dx.doi.org/10.1086/167917}{10.1086/167917}

\bibitem{Will2014} Will, C.~M., ``The Confrontation between General Relativity and Experiment,'' {\em Liv. Rev. Rel.} \textbf{17}, 4 (2014). doi:~\href{http://dx.doi.org/10.12942/lrr-2014-4}{10.12942/lrr-2014-4}
% http://adsabs.harvard.edu/abs/2014LRR....17....4W
\bibitem{Will2011} Will, C.~M., ``On the unreasonable effectiveness of the post-Newtonian approximation in gravitational physics,'' {\em PNAS} \textbf{108}, 5938--5945 (2011). doi:~\href{http://dx.doi.org/10.1073/pnas.1103127108}{10.1073/pnas.1103127108}
\bibitem{Will1986} Will, C., {\em Was Einstein Right? Putting General Relativity to the Test.} Basic Books 1986
\end{thebibliography}
\end{document}